\documentclass[prb,twocolumn]{revtex4-2}

\usepackage{textcase}
\usepackage{dcolumn}
\usepackage{bm}
\usepackage{amsmath,amsthm,amssymb,mathrsfs}
\usepackage[pdftex]{graphicx}
\begin{document}

\title{Stray magnetic fields from elliptical-shaped and stadium-shaped ferromagnets}

\author{Tomohiro Taniguchi}
\email{tomohiro-taniguchi@aist.go.jp}

\affiliation{ 
National Institute of Advanced Industrial Science and Technology (AIST), Research Center for Emerging Computing Technologies, Tsukuba 305-8568, Japan}

\date{\today}%

\begin{abstract}
An artificial spin ice consisting of numerous ferromagnets has attracted attention because of its applicability to practical devices. 
The ferromagnets interact through their stray magnetic field and show various functionality. 
The ferromagnetic element in the spin ice was recently made in elliptical-shape or stadium-shape. 
The former has a narrow edge, expecting to generate a large stray magnetic field. 
The latter has a large volume and is also expected to generate a large stray magnetic field. 
Here, we estimate the stray magnetic field by numerically integrating the solution of the Poisson equation. 
When magnetization is parallel to an easy axis, the elliptical-shaped ferromagnet generates a larger stray magnetic field than the stadium-shaped ferromagnet. 
The stray magnetic fields from both ferromagnets for arbitrary magnetization directions are also investigated. 
\end{abstract}

\maketitle


\section{Introduction}

Computation of magnetic field from a ferromagnetic body, or more generally solving a Possion equation, has been an attractive problem in both mathematics and physics \cite{osborn45,joseph65,joseph66,druyvesteyn71,chen91,aharoni98,guslienko99,tandon03,beleggia05,beleggia06,sukhostavets13,metlov13},   and still prompts publications recently \cite{taniguchi17,taniguchi18,caciagli18,almonacid18,slanovc22,masiero23}. 
Even after great development of numerical solvers \cite{nakatani89,komine95,ohe09,vansteenkiste14,tsukahara16,yoshida17}, deriving a solution of potential, or field, in terms of analytical functions or in forms of some integrals with simplified approximations, such as macrospin or rigid-vortex assumption, is highly demanded, particularly when many-body problems are of interest \cite{araujo15,taniguchi19,camsari21,castro22,yamaguchi23}. 
This is because it enables us to estimate various parameters, such as coercive and stray fields, with adequate calculation cost.
The past works have mainly focused on the internal magnetic field and derived the solutions of the demagnetization coefficients for various shapes of ferromagnets such as ellipsoid, cylinder, and cuboid  \cite{osborn45,joseph65,joseph66,druyvesteyn71,chen91,aharoni98,tandon03,beleggia05,beleggia06}, while the stray magnetic field, originated from magnetostatic interaction, for vortex, cylinder, and so on has also been investigated \cite{guslienko99,sukhostavets13,metlov13,taniguchi17,taniguchi18,caciagli18,almonacid18,slanovc22,masiero23}. 

\begin{figure}
\centerline{\includegraphics[width=1.0\columnwidth]{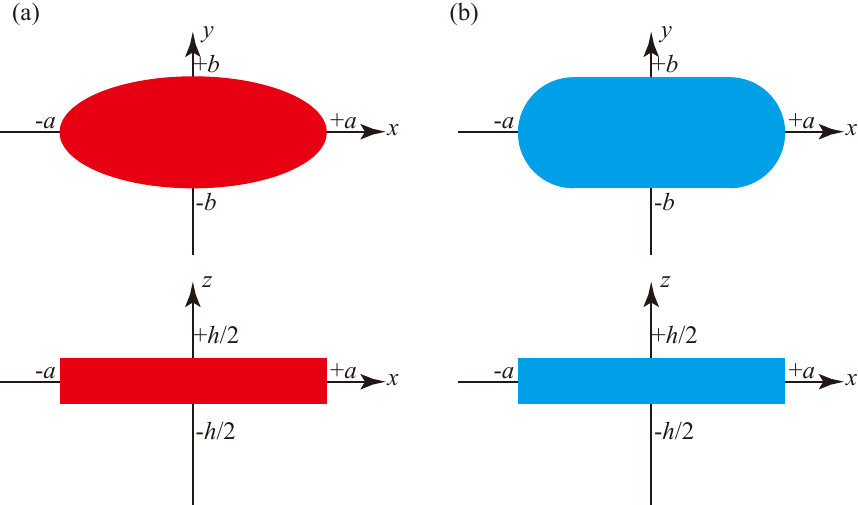}}
\caption{(Color online)
             Top and side views of (a) elliptical-shaped and (b) stadium-shaped ferromagnets. 
             The length in the long-axis ($x$-axis) direction and the width in the short-axis ($y$-axis) direction are $2a$ and $2b$, respectively, while the height in the $z$ direction is $h$. 
         \vspace{-3ex}}
\label{fig:fig1}
\end{figure}

An interesting target nowadays related to this is to compare stray magnetic fields from two kinds of uniformly magnetized ferromagnets, namely elliptical-shaped and stadium-shaped ferromagnets \cite{kubota_intermag,kubota}, which are schematically shown in Figs. \ref{fig:fig1}(a) and \ref{fig:fig1}(b), respectively. 
They are recently used as a fundamental element of artificial spin ice, and their stray magnetic fields are the origin of the frustration in it through magnetostatic interaction \cite{wang06,tanaka06,qi08,mengotti08,farhan13,farhan14,gartside18,bramwell20}. 
The frustrated states of the ferromagnets are expected to be used in several applications such as multibit memory and neuromorphic computing \cite{skjaervo20,gartside22,hu23}. 
Therefore, the evaluation and comparison from these two kinds of ferromagnets are of great interest. 
Let us assume that the length and width of these two ferromagnets, which are $2a$ and $2b$ in Fig. \ref{fig:fig1}, are the same. 
Accordingly, the stadium-shaped ferromagnet has a larger volume compared with the elliptic-shaped ferromagnet. 
From this perspective, the stadium-shaped ferromagnet is expected to generate a larger stray magnetic field. 
However, we should note that the magnetic pole of the elliptic-shaped ferromagnet is concentrated in a narrower region (around $x\simeq \pm a$ in Fig. \ref{fig:fig1}) compared with that in the stadium-shaped ferromagnet, when the magnetization points parallel to the easy (long) axis.  
In this sense, the elliptic-shaped ferromagnet is expected to generate a larger stray field. 
Then, a question arises as to which shape of the ferromagnet generates a larger stray field. 
Here, we compute their stray magnetic fields with macrospin assumption and study this question. 
Starting from Maxwell equations in a steady state, $\bm{\nabla}\cdot\mathbf{B}=0$ and $\bm{\nabla}\times\mathbf{H}=\bm{0}$, where $\mathbf{B}$ and $\mathbf{H}$ are the magnetic flux density and the magnetic field, respectively, the magnetic potential $V$, satisfying $\mathbf{H}=-\bm{\nabla}V$, obeys $\bm{\nabla}^{2}V=4\pi \bm{\nabla}\cdot\mathbf{M}$, where $\mathbf{M}$ is the magnetization. 
The solution of this Possion equation is 
\begin{equation}
  V(\mathbf{r})
  =
  \oint dS^{\prime}
  \frac{\mathbf{n}\cdot\mathbf{M}}{|\mathbf{r}-\mathbf{r}^{\prime}|}, 
  \label{eq:potential}
\end{equation}
where $dS$ is an infinitesimal cross-section area of the integral region, while $\mathbf{n}$ is a unit vector orthogonal to the surface $dS$. 
The stray magnetic field is estimated from Eq. (\ref{eq:potential}) as $\mathbf{H}=-\bm{\nabla}V$. 
It is found that the elliptical-shaped ferromagnet generates a larger stray magnetic field along the direction of the easy axis while the stadium-shaped ferromagnet generates a larger stray magnetic field along the in-plane hard-axis direction.


\section{Stray field from an elliptical-shaped ferromagnet}

\begin{figure}
\centerline{\includegraphics[width=1.0\columnwidth]{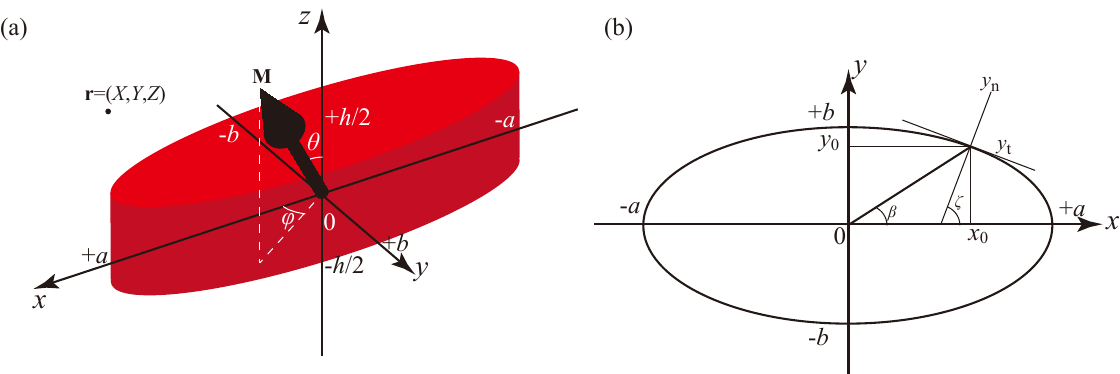}}
\caption{(Color online)
             (a) Bird's eye view of the elliptical-shaped ferromagnet. 
             (b) Top view of the elliptical-shaped ferromagnet, where the tanget $y_{\rm t}$ and normal $y_{\rm n}$ with at a point $(x_{0},y_{0})$, as well as the angles $\beta$ and $\zeta$, are indicated. 
         \vspace{-3ex}}
\label{fig:fig2}
\end{figure}

Here, we derive a formula of the stray magnetic field from an elliptical-shaped ferromagnet. 
The geometry is schematically shown in Fig. \ref{fig:fig2}(a). 
In the following, we denote that magnetization direction in terms of the zenith and azimuth angles, $\theta$ and $\varphi$, as $\mathbf{M}=M (\sin\theta\cos\varphi,\sin\theta\sin\varphi,\cos\theta)$. 
Since the Poisson equation is a linear equation, its solution is a superposition of the solutions, where the magnetization points to in-plane ($xy$-plane) and perpendicular ($z$-axis) directions. 
Therefore, we denote the solution of the magnetic potential $V$ for the in-plane and perpendicularly magnetized ferromagnets as $V_{\rm i}$ and $V_{\rm p}$, respectively. 
The measurement point of the potential is $\mathbf{r}=(X,Y,Z)$. 
The magnetic field at this point is $\mathbf{H}(\mathbf{r})=-\bm{\nabla}V|_{\mathbf{x}=\mathbf{r}}=(-\partial V/\partial x,-\partial V/\partial y,-\partial V/\partial z)_{\mathbf{x}=\mathbf{r}}$. 


\subsection{Magnetic potential}

To derive the solution of $V$ for the elliptical-shaped ferromagnet, it is useful to introduce an elliptic coordinate, where the positions $x$ and $y$ in the Cartesian coordinate are related to new variables, $\alpha$ and $\beta$, as $x=c\cosh\alpha\cos\beta$ and $y=c\sinh\alpha\sin\beta$, where $-\infty \le \alpha \le \infty$ and $-\pi \le \beta \le \pi$, while $c=\sqrt{a^{2}-b^{2}}$.  
We also introduce 
\begin{equation}
  \alpha_{\rm e}
  =
  \frac{1}{2}
  \ln
  \left(
    \frac{a+b}{a-b}
  \right). 
\end{equation}
Let us denote a point on the circumference of the ellipse as $(x_{0},y_{0})$; see also Fig. \ref{fig:fig2}(b). 
The tangent and normal with respect to the point are, respectively, given by 
\begin{equation}
  y_{\rm t}
  =
  \frac{b^{2}}{y_{0}}
  \left(
    1
    -
    \frac{x_{0}}{a^{2}}
    x
  \right),
\end{equation}
\begin{equation}
  y_{\rm n}
  =
  \frac{a^{2}y_{0}}{b^{2}x_{0}}
  x
  -
  y_{0}
  \left(
    \frac{a^{2}}{b^{2}}
    -
    1
  \right).
\end{equation}
The angle $\zeta$ between $y_{\rm n}$ and the $x$-axis satisfies 
\begin{equation}
  \cos\zeta
  =
  \frac{b^{2}\cosh\alpha_{\rm e}\cos\beta}{\sqrt{b^{4}\cosh^{2}\alpha_{\rm e}\cos^{2}\beta + a^{4}\sinh^{2}\alpha_{\rm e}\sin^{2}\beta}},
\end{equation}
\begin{equation}
  \sin\zeta
  =
  \frac{a^{2}\sinh\alpha_{\rm e}\sin\beta}{\sqrt{b^{4}\cosh^{2}\alpha_{\rm e}\cos^{2}\beta + a^{4}\sinh^{2}\alpha_{\rm e}\sin^{2}\beta}}.
\end{equation}
Therefore, the projection of the magnetization to the normal vector, $\mathbf{n}\cdot\mathbf{M}$, becomes 
$M_{\rm n}=M\sin\theta(\cos\varphi\cos\zeta+\sin\varphi\sin\zeta)$. 
An infinitesimal cross-section area around the point $(x_{0},y_{0})$ is a product of infinitesimal lines in the $xy$-plane, $\sqrt{x_{0}^{2}+y_{0}^{2}}d\beta$, and along the $z$-direction, $dz$, as $dS=\sqrt{x_{0}^{2}+y_{0}^{2}}d\beta dz=c\sqrt{\cosh^{2}\alpha_{\rm e}\cos^{2}\beta+\sinh^{2}\alpha_{\rm e}\sin^{2}\beta}d\beta dz$. 
Accordingly, the magnetic potential $V_{\rm i}$ generated from the in-plane magnetized elliptical-shaped ferromagnet becomes 
\begin{widetext}
\begin{equation}
\begin{split}
  V_{\rm i}
  =
  M & \sin\theta
  \int_{-h/2}^{h/2}
  dz^{\prime} 
  \int_{-\pi}^{\pi}
  d \beta
  \frac{c \sqrt{\cosh^{2}\alpha_{\rm e}\cos^{2}\beta+\sinh^{2}\alpha_{\rm e}\sin^{2}\beta}}
    {\sqrt{(x-c\cosh\alpha_{\rm e}\cos\beta)^{2}+(y-c\sinh\alpha_{\rm e}\sin\beta)^{2}+(z-z^{\prime})^{2}}}
\\
  &
  \times
  \left(
    \frac{b^{2}\cosh\alpha_{\rm e}\cos\beta\cos\varphi+a^{2}\sinh\alpha_{\rm e}\sin\beta\sin\varphi}{\sqrt{b^{4}\cosh^{2}\alpha_{\rm e}\cos^{2}\beta+a^{4}\sinh^{2}\alpha_{\rm e}\sin^{2}\beta}}
  \right).
  \label{eq:V_i_elliptical}
\end{split}
\end{equation}
\end{widetext}

When the magnetization points to the perpendicular ($z$) direction, the infinitesimal cross-section area is $dxdy=Jd\alpha d\beta$, where the Jacobian $J$ is $(c^{2}/2)(\cosh 2\alpha-\cos 2\beta)$. 
Note that there are two magnetic poles, $\mathbf{n}\cdot\mathbf{M}$, appearing on the $xy$-planes at $z=h/2$ and $z=-h/2$, with the opposite sign. 
Thus, the magnetic potential $V_{\rm p}$ generated from the perpendicularly magnetized elliptical-shaped ferromagnet is 
\begin{widetext}
\begin{equation}
\begin{split}
  V_{\rm p}
  =
  M & \cos\theta
  \int_{0}^{\alpha_{\rm e}} d\alpha
  \int_{-\pi}^{\pi} d\beta
  \frac{c^{2}(\cosh 2\alpha-\cos 2\beta)}{2}
\\
  &
  \left[
    \frac{1}{\sqrt{(x-c\cosh\alpha\cos\beta)^{2}+(y-c\sinh\alpha\sin\beta)^{2}+(z-h/2)^{2}}}
  \right.
\\
  &
  \left.
  -
    \frac{1}{\sqrt{(x-c\cosh\alpha\cos\beta)^{2}+(y-c\sinh\alpha\sin\beta)^{2}+(z+h/2)^{2}}}
  \right].
  \label{eq:V_p_elliptical}
\end{split}
\end{equation}
\end{widetext}


\subsection{Magnetic field}

The $x$, $y$, and $z$ components of the magnetic field are given by 
\begin{widetext}
\begin{equation}
\begin{split}
  H_{x}
  =&
  M \sin\theta
  \int_{-h/2}^{h/2}
  d z^{\prime} 
  \int_{-\pi}^{\pi}
  d \beta
  \frac{c \sqrt{\cosh^{2}\alpha_{\rm e}\cos^{2}\beta + \sinh^{2}\alpha_{\rm e}\sin^{2}\beta}(x-c \cosh\alpha_{\rm e}\cos\beta)}{[(x-c \cosh\alpha_{\rm e}\cos\beta)^{2} + (y-c \sinh\alpha_{\rm e}\sin\beta)^{2} + (z-z^{\prime})^{2}]^{3/2}}
\\
  &
  \ \ \ \ \ \ \ \ \ \times
  \left(
    \frac{b^{2}\cosh\alpha_{\rm e}\cos\beta\cos\varphi + a^{2}\sinh\alpha_{\rm e}\sin\beta\sin\varphi}{\sqrt{b^{4}\cosh^{2}\alpha_{\rm e}\cos^{2}\beta + a^{4}\sinh^{2}\alpha_{\rm e}\sin^{2}\beta}}
  \right)
\\
  &+
  M
  \cos\theta
  \int_{0}^{\alpha_{\rm e}}
  d \alpha
  \int_{-\pi}^{\pi}
  d \beta 
  \frac{c^{2}(\cosh 2\alpha-\cos 2\beta)}{2}
  \left(
    x
    -
    c \cosh\alpha
    \cos\beta
  \right)
\\
  &
  \ \ \ \ \ \times
  \left[
    \frac{1}{[(x-c \cosh\alpha\cos\beta)^{2} + (y-c \sinh\alpha\sin\beta)^{2} + (z-h/2)^{2}]^{3/2}}
  \right.
\\
  &
  \ \ \ \ \ 
  \left.
    -
    \frac{1}{[(x-c \cosh\alpha\cos\beta)^{2} + (y-c \sinh\alpha\sin\beta)^{2} + (z+h/2)^{2}]^{3/2}}
  \right],
  \label{eq:H_X_elliptical}
\end{split}
\end{equation}

\begin{equation}
\begin{split}
  H_{y}
  =&
  M \sin\theta
  \int_{-h/2}^{h/2}
  d z^{\prime} 
  \int_{-\pi}^{\pi}
  d \beta
  \frac{c \sqrt{\cosh^{2}\alpha_{\rm e}\cos^{2}\beta + \sinh^{2}\alpha_{\rm e}\sin^{2}\beta}(y-c \sinh\alpha_{\rm e}\sin\beta)}{[(x-c \cosh\alpha_{\rm e}\cos\beta)^{2} + (y-c \sinh\alpha_{\rm e}\sin\beta)^{2} + (z-z^{\prime})^{2}]^{3/2}}
\\
  &
  \ \ \ \ \ \ \ \ \ \times
  \left(
    \frac{b^{2}\cosh\alpha_{\rm e}\cos\beta\cos\varphi + a^{2}\sinh\alpha_{\rm e}\sin\beta\sin\varphi}{\sqrt{b^{4}\cosh^{2}\alpha_{\rm e}\cos^{2}\beta + a^{4}\sinh^{2}\alpha_{\rm e}\sin^{2}\beta}}
  \right)
\\
  &+
  M
  \cos\theta
  \int_{0}^{\alpha_{\rm e}}
  d \alpha
  \int_{-\pi}^{\pi}
  d \beta 
  \frac{c^{2}(\cosh 2\alpha-\cos 2\beta)}{2}
  \left(
    y
    -
    c \sinh\alpha
    \sin\beta
  \right)
\\
  &
  \ \ \ \ \ \times
  \left[
    \frac{1}{[(x-c \cosh\alpha\cos\beta)^{2} + (y-c \sinh\alpha\sin\beta)^{2} + (z-h/2)^{2}]^{3/2}}
  \right.
\\
  &
  \ \ \ \ \ 
  \left.
    -
    \frac{1}{[(x-c \cosh\alpha\cos\beta)^{2} + (y-c \sinh\alpha\sin\beta)^{2} + (z+h/2)^{2}]^{3/2}}
  \right],
  \label{eq:H_Y_elliptical}
\end{split}
\end{equation}

\begin{equation}
\begin{split}
  H_{z}
  =&
  M \sin\theta
  \int_{-h/2}^{h/2}
  d z^{\prime} 
  \int_{-\pi}^{\pi}
  d \beta
  \frac{c \sqrt{\cosh^{2}\alpha_{\rm e}\cos^{2}\beta + \sinh^{2}\alpha_{\rm e}\sin^{2}\beta}(z-z^{\prime})}{[(x-c \cosh\alpha_{\rm e}\cos\beta)^{2} + (y-c \sinh\alpha_{\rm e}\sin\beta)^{2} + (z-z^{\prime})^{2}]^{3/2}}
\\
  &
  \ \ \ \ \ \ \ \ \ \times
  \left(
    \frac{b^{2}\cosh\alpha_{\rm e}\cos\beta\cos\varphi + a^{2}\sinh\alpha_{\rm e}\sin\beta\sin\varphi}{\sqrt{b^{4}\cosh^{2}\alpha_{\rm e}\cos^{2}\beta + a^{4}\sinh^{2}\alpha_{\rm e}\sin^{2}\beta}}
  \right)
\\
  &+
  M
  \cos\theta
  \int_{0}^{\alpha_{\rm e}}
  d \alpha
  \int_{-\pi}^{\pi}
  d \beta 
  \frac{c^{2}(\cosh 2\alpha-\cos 2\beta)}{2}
\\
  &
  \ \ \ \ \ \times
  \left[
    \frac{z-h/2}{[(x-c \cosh\alpha\cos\beta)^{2} + (y-c \sinh\alpha\sin\beta)^{2} + (z-h/2)^{2}]^{3/2}}
  \right.
\\
  &
  \ \ \ \ \ 
  \left.
    -
    \frac{z+h/2}{[(x-c \cosh\alpha\cos\beta)^{2} + (y-c \sinh\alpha\sin\beta)^{2} + (z+h/2)^{2}]^{3/2}}
  \right]. 
  \label{eq:H_Z_elliptical}
\end{split}
\end{equation}
\end{widetext}


\section{Stray field from a stadium-shaped ferromagnet}

\begin{figure}
\centerline{\includegraphics[width=1.0\columnwidth]{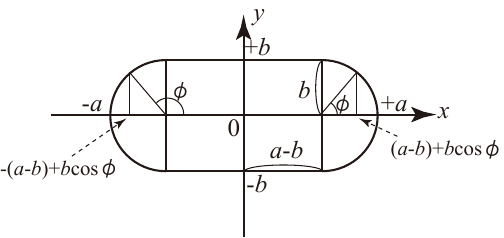}}
\caption{(Color online)
             Top view of the stadium-shaped ferromagnet, which can be regarded as a cylinder with a radius $b$ separated by a cuboid at the center. 
             Definition of $\phi$ of the cylinder coordinate is also shown. 
             The value of the $x$ coordinate of a point on the circumference is $\pm (a-b)+b \cos\phi$. 
         \vspace{-3ex}}
\label{fig:fig3}
\end{figure}

Here, we derive a formula of the stray magnetic field from a stadium-shaped ferromagnet. 
We note that the stadium-shaped ferromagnet can be regarded as a cylinder ferromagnet with a radius $b$ and height $h$, which is separated by a cuboid with a long side $2(a-b)$, short side $2b$, and height $h$; see Fig. \ref{fig:fig3}. 
Accordingly, the stray magnetic field of this ferromagnet can be obtained as a superposition of the stray magnetic fields from cylinder-shaped and cuboid-shaped ferromagnets.  
The magnetic fields generated by these ferromagnets can be expressed in terms of analytical and special functions, as shown in Supplementary Material of Ref. \cite{taniguchi17}. 
We keep, however, integral forms of the magnetic field because the expression of the field by these functions will be greatly complex. 


\subsection{Stray magnetic field generated by magnetization pointing in $x$ direction}

Let us first consider a case where the magnetization $\mathbf{M}$ points to the $x$ direction. 
The magnetic pole, $\mathbf{n}\cdot\mathbf{M}$, in this case appears only in the side of the cylinder parts. 
Thus, it is convenient to introduce a cylinder coordinate $(\rho,\phi,z)$, which relates to the Cartesian coordinate as $x=\rho\cos\phi$ and $y=\rho\sin\phi$. 
An infinitesimal cross-section area on the side of the cylinder is $b d\phi dz$, while the magnetic pole $\mathbf{n}\cdot\mathbf{M}$ on this area is $M\cos\phi$. 
Therefore, the magnetic potential is given by (see also Fig. \ref{fig:fig3})
\begin{widetext}
\begin{equation}
\begin{split}
  V_{{\rm i},x}
  =
  &
  M 
  \int_{-\pi/2}^{\pi/2}
  d \phi 
  \int_{-h/2}^{h/2}
  d z^{\prime} 
  \frac{b \cos\phi}{\sqrt{[x-(a-b)-b\cos\phi]^{2}+(y-b\sin\phi)^{2}+(z-z^{\prime})^{2}}}
\\
  &
  +  
  M 
  \int_{\pi/2}^{3\pi/2}
  d \phi 
  \int_{-h/2}^{h/2}
  d z^{\prime} 
  \frac{b \cos\phi}{\sqrt{[x+(a-b)-b\cos\phi]^{2}+(y-b\sin\phi)^{2}+(z-z^{\prime})^{2}}},  
\end{split}
\end{equation}
\end{widetext}
or by redefining $\phi$ in the second integral as $\phi-\pi$, 
\begin{widetext}
\begin{equation}
\begin{split}
  V_{{\rm i},x}
  =
  M 
  \int_{-\pi/2}^{\pi/2}
  d \phi 
  \int_{-h/2}^{h/2}
  d z^{\prime} 
&
  \left\{
    \frac{b \cos\phi}{\sqrt{[x-(a-b)-b\cos\phi]^{2}+(y-b\sin\phi)^{2}+(z-z^{\prime})^{2}}}
  \right.
\\
  &
  -
  \left.
    \frac{b \cos\phi}{\sqrt{[x+(a-b)+b\cos\phi]^{2}+(y+b\sin\phi)^{2}+(z-z^{\prime})^{2}}}
  \right\}.
\end{split}
\end{equation}
\end{widetext}
The stray magnetic field $\mathbf{H}=-\bm{\nabla}V_{{\rm i},x}$ is given by 
\begin{widetext}
\begin{equation}
\begin{split}
  H_{xx}
  =
  M 
  \int_{-\pi/2}^{\pi/2}
  d \phi 
  \int_{-h/2}^{h/2}
  d z^{\prime} 
&
  \left(
    \frac{[x-(a-b)-b\cos\phi]b\cos\phi}{\{[x-(a-b)-b\cos\phi]^{2}+(y-b\sin\phi)^{2}+(z-z^{\prime})^{2}\}^{3/2}}
  \right.
\\
  &
  -
  \left.
    \frac{[x+(a-b)+b\cos\phi]b\cos\phi}{\{[x+(a-b)+b\cos\phi]^{2}+(y+b\sin\phi)^{2}+(z-z^{\prime})^{2}\}^{3/2}}
  \right), 
\end{split}
\end{equation}
\begin{equation}
\begin{split}
  H_{yx}
  =
  M 
  \int_{-\pi/2}^{\pi/2}
  d \phi 
  \int_{-h/2}^{h/2}
  d z^{\prime} 
&
  \left(
    \frac{(y-b\sin\phi)b\cos\phi}{\{[x-(a-b)-b\cos\phi]^{2}+(y-b\sin\phi)^{2}+(z-z^{\prime})^{2}\}^{3/2}}
  \right.
\\
  &
  -
  \left.
    \frac{(y+b\cos\phi)b\cos\phi}{\{[x+(a-b)+b\cos\phi]^{2}+(y+b\sin\phi)^{2}+(z-z^{\prime})^{2}\}^{3/2}}
  \right).
\end{split}
\end{equation}
\begin{equation}
\begin{split}
  H_{zx}
  =
  M 
  \int_{-\pi/2}^{\pi/2}
  d \phi 
  \int_{-h/2}^{h/2}
  d z^{\prime} 
&
  \left(
    \frac{(z-z^{\prime})b\cos\phi}{\{[x-(a-b)-b\cos\phi]^{2}+(y-b\sin\phi)^{2}+(z-z^{\prime})^{2}\}^{3/2}}
  \right.
\\
  &
  -
  \left.
    \frac{(z-z^{\prime})b\cos\phi}{\{[x+(a-b)+b\cos\phi]^{2}+(y+b\sin\phi)^{2}+(z-z^{\prime})^{2}\}^{3/2}}
  \right).
\end{split}
\end{equation}
\end{widetext}


\subsection{Stray magnetic field generated by magnetization pointing in $y$ direction}

Next, we calculate the stray magnetic field when the magnetization points to the $y$ direction. 
The magnetic potential in this case is the sum of those generated by the magnetic poles appear on the sides of the cylinder and cuboid. 
The former contribution, similar to $V_{{\rm i},x}$ is 
\begin{widetext}
\begin{equation}
\begin{split}
  V_{{\rm i},y,{\rm cy}}
  =
  &
  M 
  \int_{-\pi/2}^{\pi/2}
  d \phi 
  \int_{-h/2}^{h/2}
  d z^{\prime} 
  \frac{b \sin\phi}{\sqrt{[x-(a-b)-b\cos\phi]^{2}+(y-b\sin\phi)^{2}+(z-z^{\prime})^{2}}}
\\
  &
  +  
  M 
  \int_{\pi/2}^{3\pi/2}
  d \phi 
  \int_{-h/2}^{h/2}
  d z^{\prime} 
  \frac{b \sin\phi}{\sqrt{[x+(a-b)-b\cos\phi]^{2}+(y-b\sin\phi)^{2}+(z-z^{\prime})^{2}}},  
\end{split}
\end{equation}
\end{widetext}
or equivalently, 
\begin{widetext}
\begin{equation}
\begin{split}
  V_{{\rm i},y,{\rm cy}}
  =
  M 
  \int_{-\pi/2}^{\pi/2}
  d \phi 
  \int_{-h/2}^{h/2}
  dz^{\prime} 
  &
  \left\{
    \frac{b \sin\phi}{\sqrt{[x-(a-b)-b\cos\phi]^{2}+(y-b\sin\phi)^{2}+(z-z^{\prime})^{2}}}
  \right.
\\
  &
  \left.
    -
    \frac{b \sin\phi}{\sqrt{[x+(a-b)+b\cos\phi]^{2}+(y+b\sin\phi)^{2}+(z-z^{\prime})^{2}}}
  \right\}, 
\end{split}
\end{equation}
while the latter contribution is 
\begin{equation}
\begin{split}
  V_{{\rm i},y,{\rm cu}}
  =
  M
  \int_{-(a-b)}^{(a-b)}
  d x^{\prime} 
  \int_{-h/2}^{h/2}
  dz^{\prime} 
  &
  \left[
    \frac{1}{\sqrt{(x-x^{\prime})^{2}+(y-b)^{2}+(z-z^{\prime})^{2}}}
    -
    \frac{1}{\sqrt{(x-x^{\prime})^{2}+(y+b)^{2}+(z-z^{\prime})^{2}}}
  \right].
\end{split}
\end{equation}
\end{widetext}
The stray magnetic field $\mathbf{H}=-\bm{\nabla}(V_{{\rm i},y,{\rm cy}}+V_{{\rm i},y,{\rm cu}})$ is given by 
\begin{widetext}
\begin{equation}
\begin{split}
  H_{xy}
  =&
  M
  \int_{-\pi/2}^{\pi/2}
  d\phi
  \int_{-h/2}^{h/2}
  dz^{\prime} 
  \left(
    \frac{[x-(a-b)-b\cos\phi]b\sin\phi}{\{[x-(a-b)-b\cos\phi]^{2}+(y-b\sin\phi)^{2}+(z-z^{\prime})^{2}\}^{3/2}}
  \right.
\\
  &\ \ \ \ \ \ \ 
  \left.
    -
    \frac{[x+(a-b)+b\cos\phi]b\sin\phi}{\{[x+(a-b)+b\cos\phi]^{2}+(y+b\sin\phi)^{2}+(z-z^{\prime})^{2}\}^{3/2}}
  \right)
\\
  &
  +
  M
  \int_{-(a-b)}^{(a-b)} 
  dx^{\prime} 
  \int_{-h/2}^{h/2}
  dz^{\prime} 
  \left\{
    \frac{x-x^{\prime}}{[(x-x^{\prime})^{2}+(y-b)^{2}+(z-z^{\prime})^{2}]^{3/2}}
    -
    \frac{x-x^{\prime}}{[(x-x^{\prime})^{2}+(y+b)^{2}+(z-z^{\prime})^{2}]^{3/2}}
  \right\},
\end{split}
\end{equation}
\begin{equation}
\begin{split}
  H_{yy}
  =&
  M
  \int_{-\pi/2}^{\pi/2}
  d\phi
  \int_{-h/2}^{h/2}
  dz^{\prime} 
  \left(
    \frac{(y-b\sin\phi)b\sin\phi}{\{[x-(a-b)-b\cos\phi]^{2}+(y-b\sin\phi)^{2}+(z-z^{\prime})^{2}\}^{3/2}}
  \right.
\\
  &\ \ \ \ \ \ \ 
  \left.
    -
    \frac{(y+b\sin\phi)b\sin\phi}{\{[x+(a-b)+b\cos\phi]^{2}+(y+b\sin\phi)^{2}+(z-z^{\prime})^{2}\}^{3/2}}
  \right)
\\
  &
  +
  M
  \int_{-(a-b)}^{(a-b)} 
  dx^{\prime} 
  \int_{-h/2}^{h/2}
  dz^{\prime} 
  \left\{
    \frac{y-b}{[(x-x^{\prime})^{2}+(y-b)^{2}+(z-z^{\prime})^{2}]^{3/2}}
    -
    \frac{y+b}{[(x-x^{\prime})^{2}+(y+b)^{2}+(z-z^{\prime})^{2}]^{3/2}}
  \right\},
\end{split}
\end{equation}
\begin{equation}
\begin{split}
  H_{zy}
  =&
  M
  \int_{-\pi/2}^{\pi/2}
  d\phi
  \int_{-h/2}^{h/2}
  dz^{\prime} 
  \left(
    \frac{(z-z^{\prime})b\sin\phi}{\{[x-(a-b)-b\cos\phi]^{2}+(y-b\sin\phi)^{2}+(z-z^{\prime})^{2}\}^{3/2}}
  \right.
\\
  &\ \ \ \ \ \ \ 
  \left.
    -
    \frac{(z-z^{\prime})b\sin\phi}{\{[x+(a-b)+b\cos\phi]^{2}+(y+b\sin\phi)^{2}+(z-z^{\prime})^{2}\}^{3/2}}
  \right)
\\
  &
  +
  M
  \int_{-(a-b)}^{(a-b)} 
  dx^{\prime} 
  \int_{-h/2}^{h/2}
  dz ^{\prime}
  \left\{
    \frac{z-z^{\prime}}{[(x-x^{\prime})^{2}+(y-b)^{2}+(z-z^{\prime})^{2}]^{3/2}}
    -
    \frac{z-z^{\prime}}{[(x-x^{\prime})^{2}+(y+b)^{2}+(z-z^{\prime})^{2}]^{3/2}}
  \right\}.
\end{split}
\end{equation}


\subsection{Stray magnetic field generated by magnetization pointing in $z$ direction}

Next, we calculate the stray magnetic field when the magnetization points to the $z$ direction. 
The magnetic pole, $\mathbf{n}\cdot\mathbf{M}$, in this case appears on the circular and rectangular surfaces on $z=\pm h/2$ with the opposite sign. 
Thus, the magnetic potential is 
\begin{equation}
\begin{split}
  V_{\rm p}
  =&
  M
  \int_{-\pi/2}^{\pi/2}
  d\phi 
  \int_{0}^{b}
  \rho d\rho
  \left\{
    \frac{1}{\sqrt{[x-(a-b)-\rho\cos\phi]^{2}+(y-\rho\sin\phi)^{2}+(z-h/2)^{2}}}
  \right.
\\
  &\ \ \ \ \ \ \ 
    -
    \frac{1}{\sqrt{[x-(a-b)-\rho\cos\phi]^{2}+(y-\rho\sin\phi)^{2}+(z+h/2)^{2}}}
\\
  &\ \ \ \ \ \ \ 
    +
    \frac{1}{\sqrt{[x+(a-b)+\rho\cos\phi]^{2}+(y+\rho\sin\phi)^{2}+(z-h/2)^{2}}}
\\
  &\ \ \ \ \ \ \ 
  \left.
    -
    \frac{1}{\sqrt{[x+(a-b)+\rho\cos\phi]^{2}+(y+\rho\sin\phi)^{2}+(z+h/2)^{2}}}
  \right\} 
\\
  &+
  M
  \int_{-(a-b)}^{(a-b)}
  dx^{\prime} 
  \int_{-b}^{b}
  dy^{\prime}
  \left[
    \frac{1}{\sqrt{(x-x^{\prime})^{2}+(y-y^{\prime})^{2}+(z-h/2)^{2}}}
  \right.
\\
  &\ \ \ \ \ \ \ 
  \left.
    -
    \frac{1}{\sqrt{(x-x^{\prime})^{2}+(y-y^{\prime})^{2}+(z+h/2)^{2}}}
  \right].
\end{split}
\end{equation}
Therefore, the stray magnetic field $\mathbf{H}=-\bm{\nabla}V_{\rm p}$ becomes 
\begin{equation}
\begin{split}
  H_{xz}
  =&
  M
  \int_{-\pi/2}^{\pi/2}
  d\phi 
  \int_{0}^{b}
  \rho d\rho
  \left(
    \frac{x-(a-b)-\rho\cos\phi}{\{[x-(a-b)-\rho\cos\phi]^{2}+(y-\rho\sin\phi)^{2}+(z-h/2)^{2}\}^{3/2}}
  \right.
\\
  &\ \ \ \ \ \ \ 
    -
    \frac{x-(a-b)-\rho\cos\phi}{\{[x-(a-b)-\rho\cos\phi]^{2}+(y-\rho\sin\phi)^{2}+(z+h/2)^{2}\}^{3/2}}
\\
  &\ \ \ \ \ \ \ 
    +
    \frac{x+(a-b)+\rho\cos\phi}{\{[x+(a-b)+\rho\cos\phi]^{2}+(y+\rho\sin\phi)^{2}+(z-h/2)^{2}\}^{3/2}}
\\
  &\ \ \ \ \ \ \ 
  \left.
    -
    \frac{x+(a-b)+\rho\cos\phi}{\{[x+(a-b)+\rho\cos\phi]^{2}+(y+\rho\sin\phi)^{2}+(z+h/2)^{2}\}^{3/2}}
  \right)
\\
  &+
  M
  \int_{-(a-b)}^{(a-b)}
  dx^{\prime} 
  \int_{-b}^{b}
  dy^{\prime}
  \left[
    \frac{x-x^{\prime}}{[(x-x^{\prime})^{2}+(y-y^{\prime})^{2}+(z-h/2)^{2}]^{3/2}}
  \right.
\\
  &\ \ \ \ \ \ \ 
  \left.
    -
    \frac{x-x^{\prime}}{[(x-x^{\prime})^{2}+(y-y^{\prime})^{2}+(z+h/2)^{2}]^{3/2}}
  \right], 
\end{split}
\end{equation}

\begin{equation}
\begin{split}
  H_{yz}
  =&
  M
  \int_{-\pi/2}^{\pi/2}
  d\phi 
  \int_{0}^{b}
  \rho d\rho
  \left(
    \frac{y-\rho\sin\phi}{\{[x-(a-b)-\rho\cos\phi]^{2}+(y-\rho\sin\phi)^{2}+(z-h/2)^{2}\}^{3/2}}
  \right.
\\
  &\ \ \ \ \ \ \ 
    -
    \frac{y-\rho\sin\phi}{\{[x-(a-b)-\rho\cos\phi]^{2}+(y-\rho\sin\phi)^{2}+(z+h/2)^{2}\}^{3/2}}
\\
  &\ \ \ \ \ \ \ 
    +
    \frac{y+\rho\sin\phi}{\{[x+(a-b)+\rho\cos\phi]^{2}+(y+\rho\sin\phi)^{2}+(z-h/2)^{2}\}^{3/2}}
\\
  &\ \ \ \ \ \ \ 
  \left.
    -
    \frac{y+\rho\sin\phi}{\{[x+(a-b)+\rho\cos\phi]^{2}+(y+\rho\sin\phi)^{2}+(z+h/2)^{2}\}^{3/2}}
  \right)
\\
  &+
  M
  \int_{-(a-b)}^{(a-b)}
  dx^{\prime} 
  \int_{-b}^{b}
  dy^{\prime}
  \left[
    \frac{y-y^{\prime}}{[(x-x^{\prime})^{2}+(y-y^{\prime})^{2}+(z-h/2)^{2}]^{3/2}}
  \right.
\\
  &\ \ \ \ \ \ \ 
  \left.
    -
    \frac{y-y^{\prime}}{[(x-x^{\prime})^{2}+(y-y^{\prime})^{2}+(z+h/2)^{2}]^{3/2}}
  \right], 
\end{split}
\end{equation}
\begin{equation}
\begin{split}
  H_{zz}
  =&
  M
  \int_{-\pi/2}^{\pi/2}
  d\phi 
  \int_{0}^{b}
  \rho d\rho
  \left(
    \frac{z-h/2}{\{[x-(a-b)-\rho\cos\phi]^{2}+(y-\rho\sin\phi)^{2}+(z-h/2)^{2}\}^{3/2}}
  \right.
\\
  &\ \ \ \ \ \ \ 
    -
    \frac{z+h/2}{\{[x-(a-b)-\rho\cos\phi]^{2}+(y-\rho\sin\phi)^{2}+(z+h/2)^{2}\}^{3/2}}
\\
  &\ \ \ \ \ \ \ 
    +
    \frac{z-h/2}{\{[x+(a-b)+\rho\cos\phi]^{2}+(y+\rho\sin\phi)^{2}+(z-h/2)^{2}\}^{3/2}}
\\
  &\ \ \ \ \ \ \ 
  \left.
    -
    \frac{z+h/2}{\{[x+(a-b)+\rho\cos\phi]^{2}+(y+\rho\sin\phi)^{2}+(z+h/2)^{2}\}^{3/2}}
  \right) 
\\
  &+
  M
  \int_{-(a-b)}^{(a-b)}
  dx^{\prime} 
  \int_{-b}^{b}
  dy^{\prime}
  \left[
    \frac{z-h/2}{[(x-x^{\prime})^{2}+(y-y^{\prime})^{2}+(z-h/2)^{2}]^{3/2}}
  \right.
\\
  &\ \ \ \ \ \ \ 
  \left.
    -
    \frac{z+h/2}{[(x-x^{\prime})^{2}+(y-y^{\prime})^{2}+(z+h/2)^{2}]^{3/2}}
  \right].
\end{split}
\end{equation}
\end{widetext}


\subsection{Total stray magnetic field generated from a stadium-shaped ferromagnet}

Summarizing the results above, the stray magnetic field from the stadium-shaped ferromagnet, when the magnetization points to an arbitrary direction as $\mathbf{M}=M(\sin\theta\cos\varphi,\sin\theta\sin\varphi,\cos\theta)$, is 
\begin{equation}
  \mathbf{H}
  =
  \begin{pmatrix}
    H_{xx} & H_{xy} & H_{xz} \\
    H_{yx} & H_{yy} & H_{yz} \\
    H_{zx} & H_{zy} & H_{zz}
  \end{pmatrix}
  \begin{pmatrix}
    \sin\theta\cos\varphi \\
    \sin\theta\sin\varphi \\
    \cos\theta 
  \end{pmatrix}
\end{equation}



\section{Computation of magnetic field}

Here, we examine the stray magnetic fields estimated by using the above formulas. 
Before that, however, let us mention the validity of the above calculation briefly. 
We used the macrospin assumption in the derivation of the stray magnetic field. 
Let us also assume there is no bulk nor interfacial magnetic anisotropy. 
Thus, the magnetic anisotropy is solely determined by the shape magnetic anisotropy.  
Then, the $x$ axis is the easy axis, while the $y$ axis is the in-plane hard axis. 
We performed micromagnetic simulation by MuMax3 \cite{vansteenkiste14} and found that, in the absence of an external magnetic field, the macrospin assumption works well when the magnetization points to the $x$ direction, while it does not work for the magnetization pointing in the $y$ direction; see Appendix. 
We, however, evaluate the stray magnetic field with the macrospin assumption even when the magnetization direction is not parallel to the $x$ direction. 
This is because, in some practical applications such as physical reservoir computing by an artificial spin ice, a large external magnetic field is applied to the whole system \cite{hon21}. 
In such a case, the macrospin assumption works even when the magnetization direction is deviated from the $x$ (easy) axis; see also Appendix. 
Therefore, we assume in the following that an external magnetic field is applied to the $x$, $y$, or $z$ direction to make the macrospin assumption used in the above calculations applicable. 
A further comparison with micromagnetic simulation is kept as a future work because it strongly depends on the system size, and its comprehensive study is beyond the scope of this work. 


\subsection{Magnetic fields when magnetization points to $x$, $y$, or $z$ directions}

Here, we show the stray magnetic fields from the elliptical-shaped and stadium-shaped ferromagents, where the magnetization is parallel to the $x$, $y$, or $z$ axis. 
The values of the parameters are $M=1500$ emu/cm${}^{3}$, $a=200$ nm, $b=75$ nm, $h=20$ nm \cite{kubota_intermag,kubota}. 
We compute the stray magnetic fields in a plane parallel to the $xy$-plane at $z=15$ nm, which is $5$ nm above the upper ferromagnetic surface at $z=+h/2=10$ nm. 


\begin{figure*}
\centerline{\includegraphics[width=2.0\columnwidth]{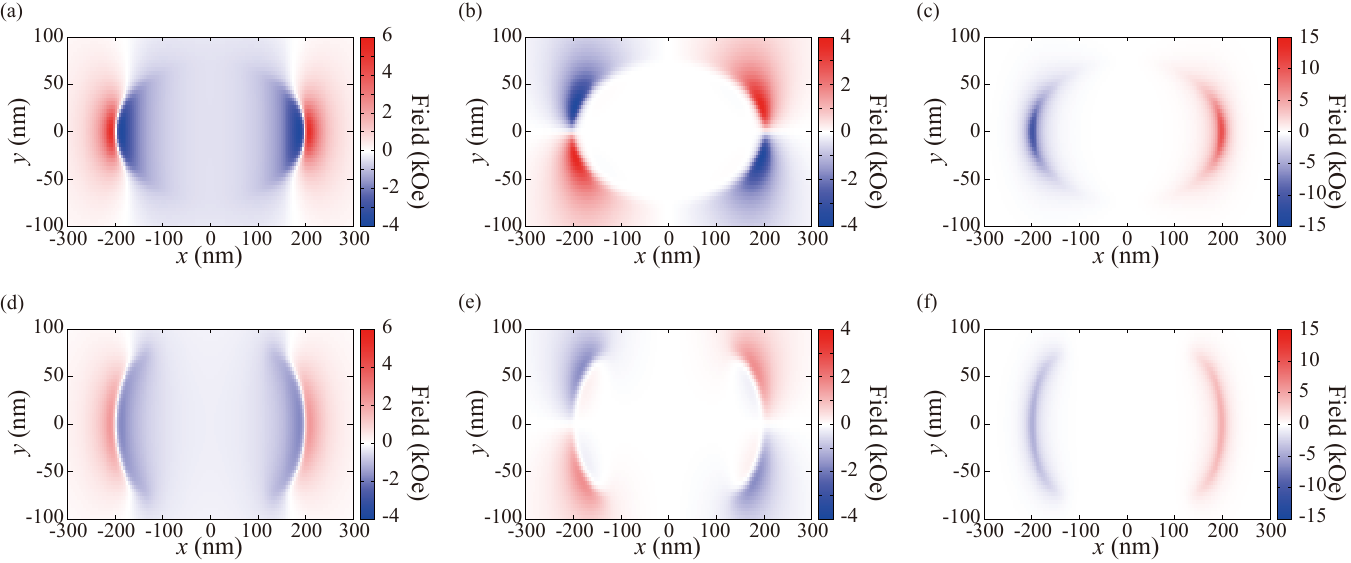}}
\caption{(Color online)
         Stray fields in (a) $x$, (b) $y$, and (c) $z$ directions generated from an elliptical-shaped ferromagnet. 
         The magnetization points to the $+x$ direction. 
         Those fields generated from a stadium-shaped ferromagent are shown in (d), (e), and (f). 
         \vspace{-3ex}}
\label{fig:fig4}
\end{figure*}


Let us first consider the case where the magnetization is parallel to the easy ($x$) axis. 
Figures \ref{fig:fig4}(a)-\ref{fig:fig4}(c) are the $x$, $y$, and $z$ components of the stray magnetic fields generated from the elliptical-shaped ferromagnet, while Figs. \ref{fig:fig4}(d)-\ref{fig:fig4}(f) show those generated from the stadium-shaped ferromagnet. 
Since the magnetization points to the $+x$ direction, the $x$ component of the stray magnetic field points to the $+x$ direction for $|x|>a$, while it points to the $-x$ direction in $|x|<a$ because the magnetic field emitted near $x=+a$ returns near to $x=-a$; see Figs. \ref{fig:fig4}(a) and \ref{fig:fig4}(d). 
The $y$ and $z$ components of the stray magnetic fields in Figs. \ref{fig:fig4}(b), \ref{fig:fig4}(c), \ref{fig:fig4}(e), and \ref{fig:fig4}(f) can also be explained in a similar way. 


Now let us remind our motivation to this study here. 
The stadium-shaped ferromagnet has a larger volume than the elliptical-shaped ferromagnet. 
In fact, their volumes are $\mathcal{V}_{\rm s}=[\pi b^{2}+4b(a-b)]h$ and $\mathcal{V}_{\rm e}=\pi abh$, and thus, their difference satisfies $\mathcal{V}_{\rm s}-\mathcal{V}_{\rm e}=(4-\pi)b(a-b)h>0$.  
Since the stray magnetic field is, roughly speaking, total number of the magnetic moments, which is the product of the magnetization and the volume, the stadium-shaped ferromagnet is expected to produce a larger stray field. 
The elliptical-shaped ferromagnet, however, has a narrower area near $|x|\simeq a$, and thus, the density of the magnetic pole is larger than that of the stadium-shaped ferromagnet. 
Therefore, it is unclear which type of the ferromagnet can generate a larger stray field. 
We thus evaluate the stray magnetic field along the $x$ direction and notice that the stray magnetic fields generated from the elliptical-shaped ferromagnets are mainly larger than those generated from the stadium-shaped ferromagnet. 
For example, the stray magnetic field generated from the elliptical-shaped ferromagnet at $x=-300$ nm and $y=0$ nm, i.e., $100$ nm external to the ferromagnetic edge at $x=-a=-200$ nm, is $441$ Oe, while that generated from the stadium-shaped ferromagnet is $281$ Oe.


\begin{figure*}
\centerline{\includegraphics[width=2.0\columnwidth]{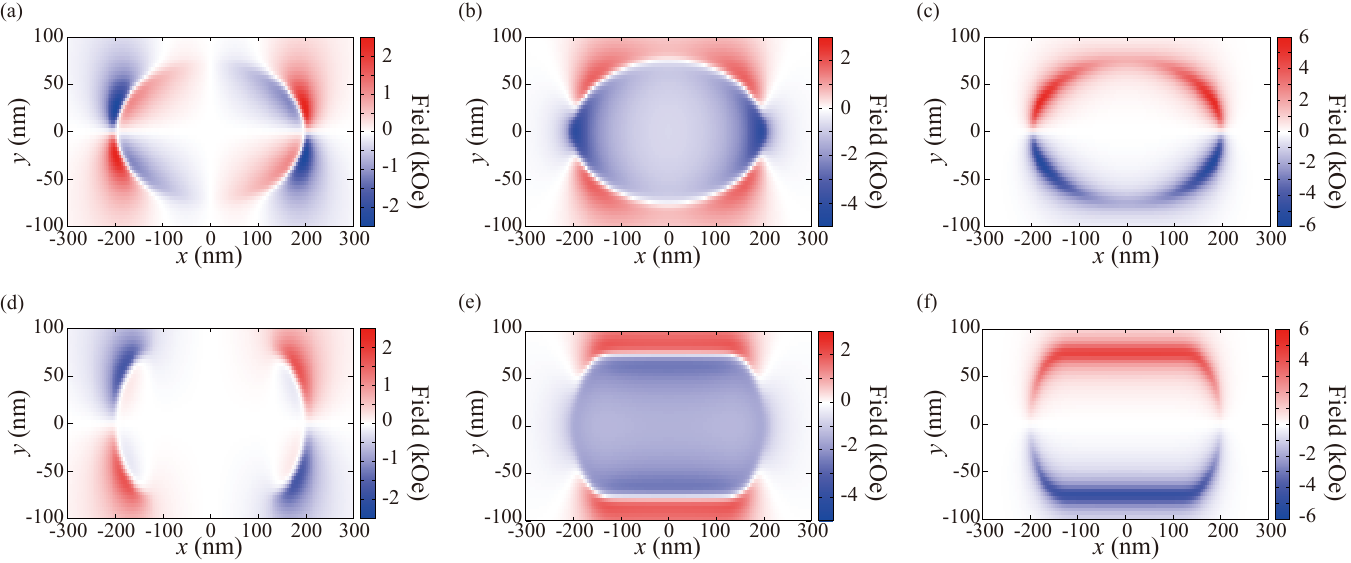}}
\caption{(Color online)
         Stray fields in (a) $x$, (b) $y$, and (c) $z$ directions generated from an elliptical-shaped ferromagnets. 
         The magnetization points to the $+y$ direction. 
         Those fields generated from a stadium-shaped ferromagents are shown in (d), (e), and (f). 
         \vspace{-3ex}}
\label{fig:fig5}
\end{figure*}


The situation is changed when the magnetization points to the $y$ direction. 
Figures \ref{fig:fig5}(a)-\ref{fig:fig5}(c) show the $x$, $y$, and $z$ components of the stray magnetic field generated from the elliptical-shaped ferromagnet, where the magnetization points to the $y$ direction. 
Those generated from the stadium-shaped ferromagnet is shown in Figs. \ref{fig:fig5}(d)-\ref{fig:fig5}(f). 
The field distribution is, roughly speaking, rotated $90^{\circ}$ from those shown in Fig. \ref{fig:fig4}. 
In particular, let us focus on Figs. \ref{fig:fig5}(b) and \ref{fig:fig5}(e). 
We notice that the $y$ component of the stray field is larger for the stadium-shaped ferromagnet that that for the elliptical-shaped ferromagnet. 
For example, the stray magnetic field at $x=0$ nm and $y=100$ nm, i.e., $25$ nm external to the ferromagnetic edge at $y=+b=75$ nm, is $1498$ Oe for the stadium-shaped ferromagnet, while that generated from the elliptical-shaped ferromagnet is $597$ Oe. 
The large difference may arise from the difference of the ferromagnetic boundary. 
The magnetic field becomes large when the surface density of the magnetic pole, given by $\mathbf{M}\cdot\mathbf{n}dS$, is large. 
For the stadium-shaped ferromagnet, the boundary of  the ferromagnet in $-(a-b)\le x \le +(a-b)$ is parallel to the $x$ axis, i.e., the vector $\mathbf{n}$, which is normal to the area $dS$, is parallel to the magnetization. 
Thus, the density of the magnetic pole, as well as the stray magnetic field, is large. 
On the other hand, for the elliptical-shaped ferromagnet, the boundary is bent, and thus the vector $\mathbf{n}$ is not parallel expect at $x=0$. 
Thus, the magnetic pole density, $\mathbf{M}\cdot\mathbf{n}dS$, becomes relatively small, resulting in a small stray magnetic field in the $y$ direction.



\begin{figure*}
\centerline{\includegraphics[width=2.0\columnwidth]{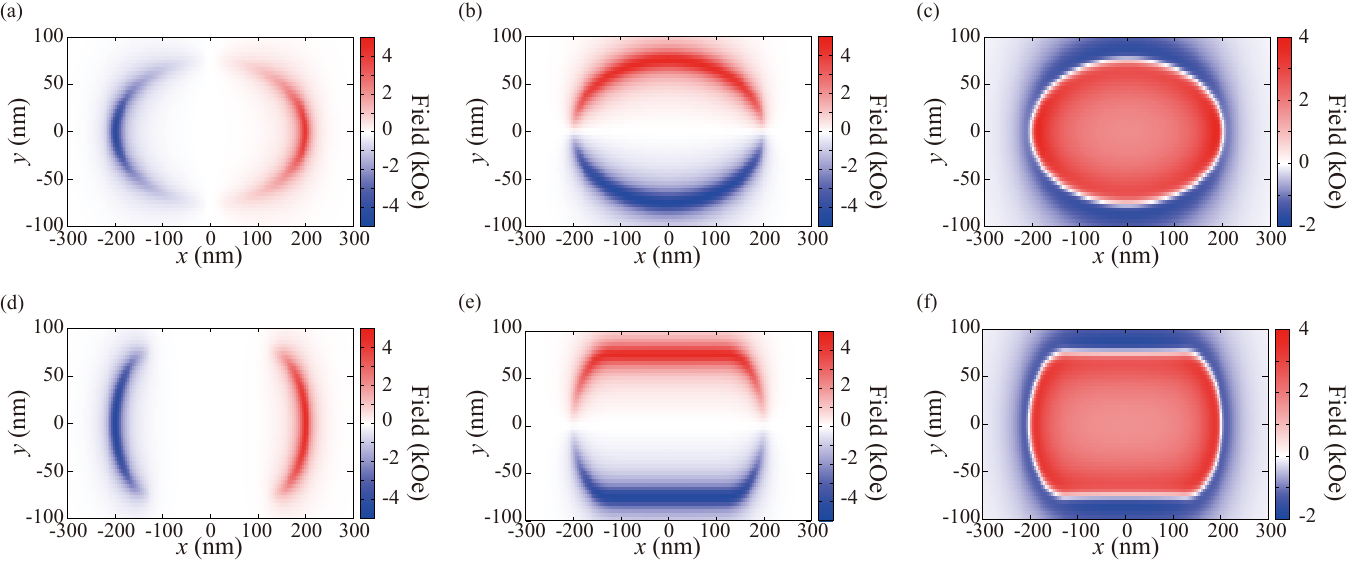}}
\caption{(Color online)
         Stray fields in (a) $x$, (b) $y$, and (c) $z$ directions generated from an elliptical-shaped ferromagnets. 
         The magnetization points to the $+z$ direction. 
         Those fields generated from a stadium-shaped ferromagents are shown in (d), (e), and (f). 
         \vspace{-3ex}}
\label{fig:fig6}
\end{figure*}


When the magnetization points to the $z$ direction, as shown in Fig. \ref{fig:fig6}, the stray magnetic fields from the elliptical-shaped and stadium-shaped ferromagnets have similar strength, although their distributions are different depending on the sample shape. 



\begin{figure}
\centerline{\includegraphics[width=1.0\columnwidth]{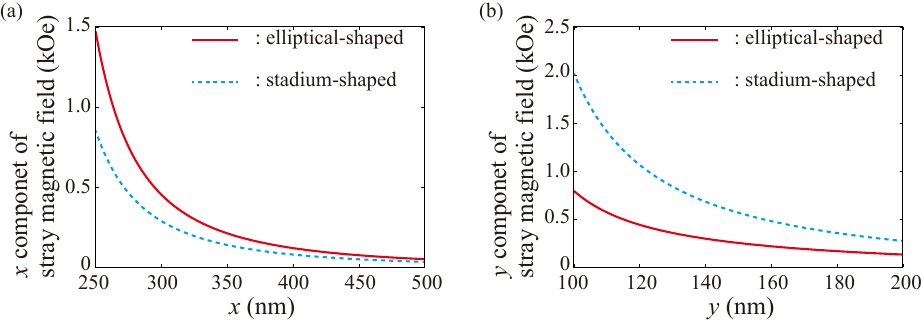}}
\caption{(Color online)
         (a) The $x$ components of the stray magnetic fields generated from the elliptical-shaped (red-solid) and the stadium-shaped (blue-dotted) ferromagnets, where the magnetization points to the $+x$ direction. 
             The measurement points locate at $y=z=0$. 
         (a) The $y$ components of the stray magnetic fields generated from the elliptical-shaped (solid red) and the stadium-shaped (dotted blue) ferromagnets, where the magnetization points to the $+y$ direction. 
             The measurement points locate at $x=z=0$.  
         \vspace{-3ex}}
\label{fig:fig7}
\end{figure}


In Fig. \ref{fig:fig7}(a), the dependence of the $x$ components of the stray magnetic fields generated from the elliptical-shaped and the stadium-shaped ferromagnets on the distance from the sample edge at $x=+a=200$ nm are shown by the red-solid and blue-dotted lines, respectively.
The $y$ and $z$ coordinates are $0$ nm, i.e., Fig. \ref{fig:fig7}(a) shows the stray magnetic field along the $x$ axis, when the magnetization points to $+x$ direction. 
It is shown that the stray magnetic field generated from the elliptical-shaped ferromagnet is larger than that generated from the stadium-shaped ferromagnet, as implied from Figs. \ref{fig:fig4}(a) and \ref{fig:fig4}(d). 
Similarly, Fig. \ref{fig:fig7}(b) shows the dependence of the $y$ component of the stray magnetic field from both ferromagnets on the distance from the sample edge at $y=+b=75$ nm, where $x=z=0$ nm.
The magnetization in this case is aligned in $+y$ direction. 
It can be confirmed that the stray magnetic field generated from the stadium-shaped ferromagnet is larger than that generated from the elliptical-shaped ferromagnet, as mentioned above. 

In previous studies \cite{wang06,tanaka06,qi08,mengotti08,farhan13,farhan14,gartside18} on the artificial spin ice consisting of ferromagnets, the magnetization alignments in equilibrium were on interest, where the magnetization in each ferromagnet approximately points to the easy-axis direction, i.e., $+x$ or $-x$ direction. 
The frustration caused by the stray magnetic field results in a large number of degenerated ground state. 
It provides several interesting physical phenomena, such as residual entropy and spin liquid. 
The presence of a large number of the degenerated ground state is also of interest from a viewpoint of practical applications such as multibit memory and logic device \cite{skjaervo20}. 
In this sense, the elliptical-shaped ferromagnet will be preferable, compared with the stadium-shaped ferromagnet, because it can generate a larger stray field in the $x$ direction and drives a strong frustration. 
In addition, the stadium-shaped ferromagnet sometimes includes vortexes at the edges, depending on the size, and breaks the macrospin assumption  \cite{kubota_intermag,kubota}. 
It results in a disappearance of magnetic pole, which is a topological charge in a spin ice \cite{skjaervo20} and plays a central role in frustration. 
This is another reason preferring the elliptical-shaped ferromanget, rather than the stadium-shaped ferromagnet, applied to an artificial spin ice. 
The stadium-shaped ferromagnet might be, however, of use to different purposes, such as reservoir computing utilizing a ferromagnetic ensemble \cite{gartside22,hu23,hon21}, where an external magnetic field is applied to the direction deviated from the easy ($x$) axis direction. 
In such a case, the stray magnetic field, pointing in a direction different from the $x$-axis, may be generated and determines the magnetization alignment of the other ferromagnets. 
Then, the history of the applied field can be recognized from the magnetization alignment.  
The shape of the ferromagnet should be designed, depending on the purpose, and large-scale simulations, as well as experiments, will be necessary in future.



\section{Conclusion}

In summary, the stray magnetic fields generated from an elliptical-shaped and stadium-shaped ferromagnets were evaluated from the integral forms of the solutions of the Poisson equation. 
The results here will be useful for the estimation of the stray magnetic field with low calculation cost, as well as relatively high reliability than a point-particle approximation \cite{hon21}. 
The elliptical-shaped ferromagnet generates a larger stray field than the stadium-shaped ferromagnet, when the magnetization points to the easy (long) axis direction. 
This is due to a larger density of the magnetic pole concentrated near the sample edge. 
The stadium-shaped ferromagnet, on the other hand, generates a larger stray magnetic field than the elliptical-shaped ferromagnet along the in-plane hard (short) axis direction because of the large number of the magnetic pole generated on the boundary. 
Regarding these results, the elliptical-shaped ferromagnet will be suitable when used in, for example, an artificial spin ice utilizing a large number of degenerated ground state. 
This is because, in such a system, the stray magnetic field generated by the magnetization which is approximately parallel to the easy axis causes a frustration and provides functionalities for multibit memory and so on. 
The stadium-shaped ferromagnet might be, on the other hand, of interest for reservoir computing utilizing ferromagnetic ensemble. 
This is because the magnetization direction is forcibly deviated from the easy axis by an external magnetic field, and thus, the stray magnetic field pointing in an arbitrary direction affects the magnetization alignment of the other ferromagnets.


\section*{Acknowledgement}

The author is grateful to Takehiko Yorozu for his enormous contribution to this work. 
The author is also thankful to Hitoshi Kubota for discussion. 
This work was supported by a JSPS KAKENHI Grant, Number 20H05655.


\appendix


\section{Verification of macrospin assumption}


The applicability of the macrospin assumption  was studied by performing micromagnetic simulation with MuMax3 \cite{vansteenkiste14}. 
The elliptical-shaped ferromagnet with $a=200$ nm, $b=75$ nm, and $h=20$ nm was used for this purpose. 
The mesh number is $200$, $75$, and $10$ in the $x$, $y$, and $z$ direction, respectively, where we remind that the sample length (width) in the $x$ ($y$) direction is $2a$ ($2b$), and thus, the mesh scale is $2$ nm in every direction. 
The exchange stiffness is assumed to be $1.3\times 10^{-6}$ erg/cm.


\begin{figure}
\centerline{\includegraphics[width=1.0\columnwidth]{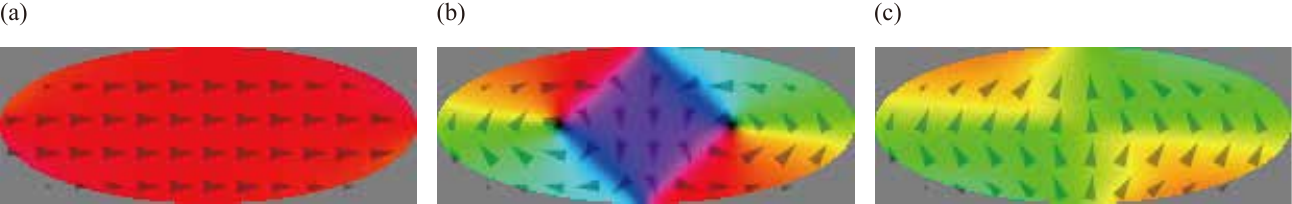}}
\caption{(Color online)
         Magnetization alignment of an elliptical-shaped ferromagnet, in the absence of an external magnetic field, when the initial state is parallel to (a) $x$ and (b) $y$ axis. 
         (c) Magnetization alignment in the presence of an external magnetic field of $1.0$ kOe in the $y$ direction, where the initial state is parallel to the $y$ axis. 
         \vspace{-3ex}}
\label{fig:fig8}
\end{figure}


Figure \ref{fig:fig8}(a) shows the magnetization alignment after relaxation, where the initial state of the magnetization was set to be uniform in the $x$ direction. 
An external magnetic field is absence in this example. 
In this case, even after the relaxation, the magnetization approximately points to the $x$ direction uniformly. 
When the initial state was set to be uniform in the $y$ direction, on the other hand, the magnetization alignment after relaxation becomes nonuniform, as schematically shown in Fig. \ref{fig:fig8}(b). 
In this case, the magnetization state is energetically unstable, and the magnetic moments start to change their direction to minimize the magnetic energy. 
The magnetization moments near the edge try to change their direction along the tangent of the ferromagnet, and as a result, magnetic vortexes appear. 
In this case, the macrospin assumption does not work well. 
The magnetic vortexes will also appear near the edges of the stadium-shaped ferromagnet due to its circular shape \cite{kubota_intermag,kubota}. 
When an external magnetic field is applied in the $y$ direction, however, a uniform alignment of the magnetization is approximately recovered, as shown in Fig. \ref{fig:fig8}(c), where the initial state was uniform to the $y$ direction and the magnitude of the external magnetic field is $1.0$ kOe.
Note that, for $a=200$ nm, $b=75$ nm, and $h=20$ nm, the demagnetization coefficients \cite{beleggia05} are $N_{x}=0.0406\cdots$, $N_{y}=0.1549\cdots$, and $N_{z}=0.8045\cdots$, and thus, the in-plane magnetic anisotropy field with $M=1500$ is $4\pi M (N_{y}-N_{x})=2.153$ kOe.



%



\end{document}